\documentclass[pre,twocolumn,superscriptaddress,floatfix,nofootinbib]{revtex4-1}  

\usepackage[paperwidth=210mm,paperheight=297mm,centering,hmargin=2cm,vmargin=2.5cm]{geometry}

\usepackage[pdftex]{graphicx}
\usepackage{amsmath,bm}
\usepackage{natbib}
\usepackage{MnSymbol}
\usepackage{csquotes}
\usepackage{mathtools}

\usepackage{hyperref}
\usepackage{xcolor} 
\usepackage{array}
\usepackage{pgfplots}
\usepackage{graphicx}
\usepackage[left,modulo,pagewise]{lineno}
\usepackage{dsfont}

\definecolor{bluemoi}{rgb}{0.25,0.50 ,0.75} 

\hypersetup{
  bookmarksopen=true,
  pdftitle=" ",
  pdfauthor=" ", 
  pdfsubject=" ", 
  pdftoolbar=true, 
  pdfmenubar=true, 
  pdfhighlight=/O, 
  colorlinks=true, 
  pdfpagemode=UseNone, 
  pdfpagelayout=SinglePage, 
  pdffitwindow=true, 
  linkcolor=bluemoi, 
  citecolor=bluemoi, 
  urlcolor=bluemoi 
}

\makeatletter
\renewcommand{\figurename}{\sf \textbf{Figure}}
\renewcommand{\thefigure}{\arabic{figure}}
\renewcommand{\fnum@figure}{\sf\textbf{\figurename}~\textbf{\thefigure}}
\renewcommand{\tablename}{\sf\textbf{Table}}
\renewcommand{\thetable}{\arabic{table}}
\renewcommand{\fnum@table}{\sf\textbf{\tablename}~\textbf{\thetable}}
\makeatother

\begin{document}

\title{Mapping mobile service usage diversity in cities}

\author{Maxime Lenormand}
\thanks{Corresponding authors: maxime.lenormand@inrae.fr}
\affiliation{TETIS, Univ Montpellier, AgroParisTech, Cirad, CNRS, INRAE, Montpellier, France}

\begin{abstract} 
The ubiquitous use of mobile devices and associated Internet services generates vast volumes of geolocated data, offering valuable insights into human behaviors and their interactions with urban environments. Over the past decade, mobile phone data have proven indispensable in various fields such as demography, geography, transport planning, and epidemiology. They enable researchers to examine human mobility patterns on unprecedented scales and analyze the spatial structure and function of cities. The relationship between mobile phone data and land use has also been extensively explored, particularly in inferring land use patterns from spatiotemporal activity. However, many studies rely on Call Detail Records (CDR) or eXtended Detail Records (XDR), which may not capture specific mobile application usage. This study aims to address this gap by mapping mobile service usage diversity in 20 French cities and investigating its correlation with land use distribution. Utilizing a Shannon diversity index, the study evaluates mobile service usage diversity based on hourly traffic volume data from 17 mobile services. Furthermore, the study compares temporal diversity with land distribution both within and among cities.
\end{abstract}

\maketitle

\section*{Introduction}

The growing and widespread use of mobile devices and associated Internet services generate daily a large volume of geolocated data. This avalanche of data has allowed us to better understand human behaviours and their links with the city's structure and function \citep{Lenormand2016}. Indeed, in the last decade, mobile phone data have proved useful in numerous applications in demography \citep{Deville2014}, geography \citep{Cottineau2019}, transport planning \citep{Jiang2016} and the spread of epidemics \citep{Tizzoni2014,Finger2016}, to cite but a few examples.

In particalar, they have allowed researchers to study individual human mobility patterns at an unprecedented scale \citep{Gonzalez2008,Blondel2015,Barbosa2018}. In addition, these data can be aggregated in space and time to analyze the city’s spatial structure \citep{Louail2014} and function \citep{Lenormand2015}. The link between mobile phone data and land use has also been widely studied in the last ten years. A particular focus has been directed towards land use inference based on the spatio-temporal activity of mobile phone users \citep{Soto2011,Toole2012,Pei2014,Lenormand2015}. Several studies have also analysed mobile phone usage patterns in relation to land use characteristics \citep{Bernini2019,Yang2019,Novovic2020} and found interesting correlations between information derived from both data sources.

However, these works usually rely on Call Detail Records (CDR) or eXtended Detail Records (XDR). They generally provide information on the distribution of mobile phone users in space (cell phone tower resolution) and time (to the second). CDR data is generated if and only if a user makes or receives a phone call. This is not the case with XDR data, where a user's location is recorded each time the user and/or device initiates download/upload operation on the Internet. In both cases the user-device(-Internet) interaction provides information about the type of signal sent, but rarely about the consumption of specific mobile applications which is fast becoming a key element in the study of human behaviours \citep{Marquez2017,Singh2019,Okic2019,Martinez2023,Goel2023}. Indeed, the analysis of high spatio-temporal resolution usage of specific mobile services is a relatively new research direction \citep{Marquez2017}, primarily constrained by the scarcity of available datasets. This type of information have been recently used to analyse the time and space characteristics of mobile service usage \citep{Marquez2017,Okic2019} or for the prediction of socio-economic indicators \citep{Goel2023}. They have notably allowed to highlight differences between urban and rural area in France \citep{Singh2019}. This new source of information has recently been used to predict socio-economic indicators \citep{Goel2023} to analyse the time and space characteristics of mobile service usage \citep{Marquez2017,Okic2019}, highlighting differences between urban and rural usage in France \citep{Singh2019}.

The purpose of this study is to map the diversity of mobile service usage in cities and explore its relationship with land use distribution. Specifically, this work has two objectives: to assess the diversity of mobile service usage in 20 French cities using a Shannon diversity index, and to focus on the hourly traffic volume information of 17 mobile services. Our focus was on examining how diversity changes over time and highlighting differences between cities based on their geographical locations within the map of France. Our second objective is to compare the temporal diversity and land distribution within and between cities. Following the approach proposed in \citep{Bernini2019}, we clustered part of the cities showing similar diversity in mobile phone usage over time and land use features separately, and then assessed the agreement between the clustering results derived from these two distinct data sources.

\section*{Materials and Methods}

\subsection*{Mobile phone data and diversity metric}

\begin{figure}[!h]
	\centering 
	\includegraphics[width=\linewidth]{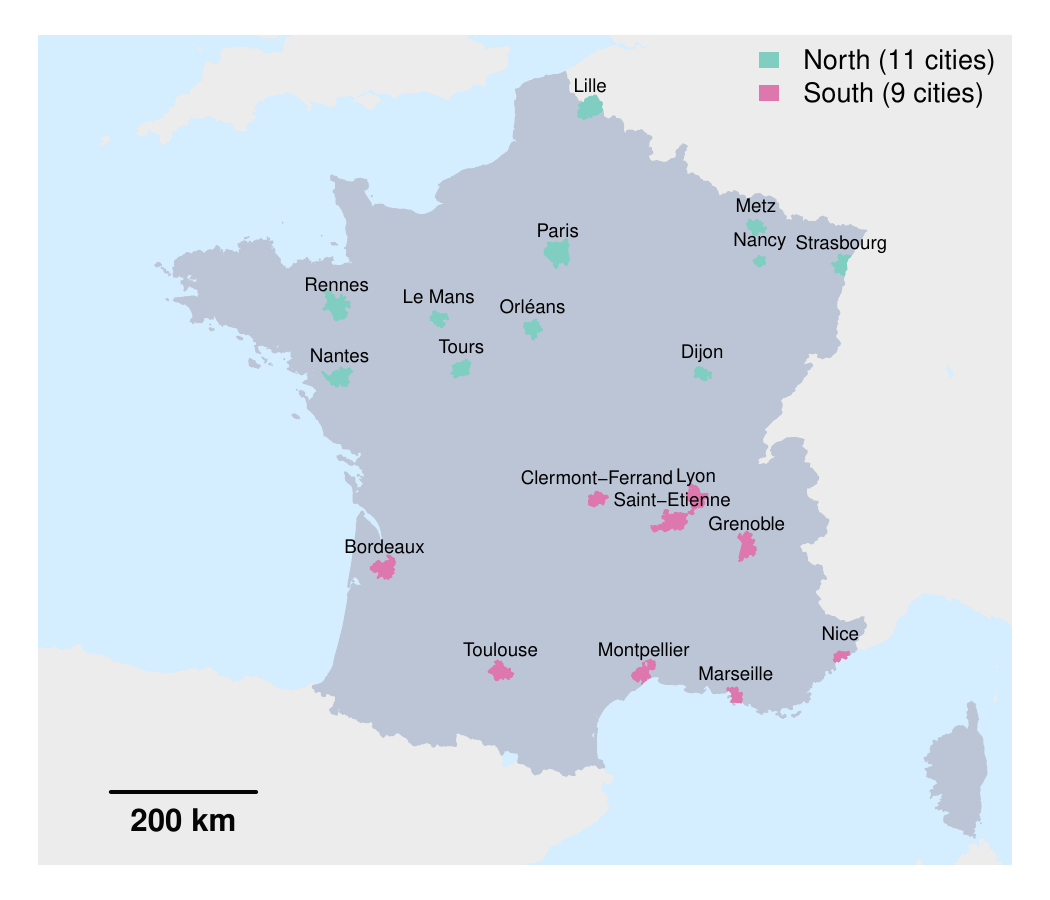}
	\caption{\textbf{Map of the studied areas.} The dataset comprises 20 French cities, with 11 northern cities marked in green and 9 southern cities marked in red.} 
	\label{Fig1}
\end{figure}

The dataset used in this study is part of the NetMob 2023 Data Challenge \citep{Martinez2023}. It contains download and upload traffic volume information on 68 mobile services during 77 days (from March 16 to May 31 2019) accross 20 cities in France (Figure \ref{Fig1}) at a 100 x 100 m$^2$ spatial resolution with a 15 minute temporal resolution.

To illustrate our approach we selected 17 mobile services among the 68 mobile services available in the dataset. These 17 mobile services have been selected based on the type of services and traffic volume (see Figure S1 in Appendix for more details). We selected and aggregated the traffic volumes for each city over three weeks (March 18-24, April 1-7 and May 13-19). We have selected these three weeks to exclude school holidays, national holidays, and any irregularities. 

\begin{figure*}
	\centering 
	\includegraphics[width=\linewidth]{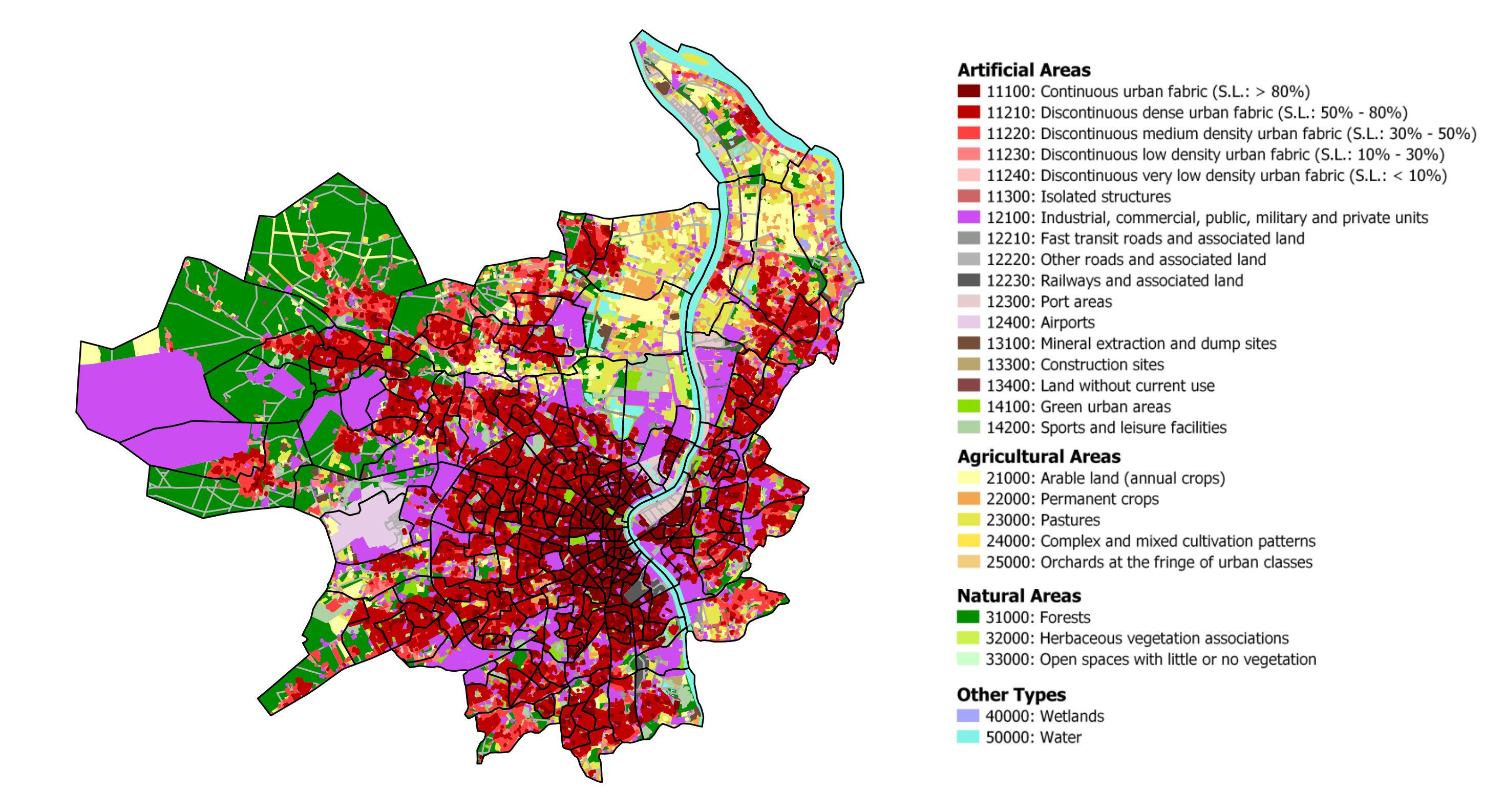}
	\caption{\textbf{Land use distribution in the city of Bordeaux.} The black lines represent the IRIS's boundaries.} 
	\label{Fig2}
\end{figure*}

The dataset was aggregated at the IRIS resolution. IRIS stands for \enquote{aggregated units for statistical information} in French and is used by the French National Institute of Statistics to disseminate infra-municipal data. The 20 cities were divided into $6,906$ IRIS based on the IRIS boundaries in 2021\footnote{\url{https://geoservices.ign.fr/contoursiris}}. Table \ref{Tab1} presents the surface area, number of IRIS and average IRIS surface area per city. An example of IRIS division in the city of Bordeaux is displayed in Figure \ref{Fig2}. The dataset was aggregated in time using 1 hour-time slots instead of the original 15 minutes slots. We only considered four days of the week: Thursday representing a normal working day, Friday, Saturday and Sunday. 

\begin{table}
	\caption{\textbf{Surface area (in km$^2$), number of IRIS and average IRIS surface area (in km$^2$) per city.}}
	\label{Tab1}
	\centering
	\vspace{0.2cm}
	\begin{tabular}{lccc}	
		
		\hline
		\\[-0.9em]
		\textbf{City} & \textbf{Area} & \textbf{IRIS} & \textbf{Area IRIS} \\ 
		\\[-0.9em]		
		\hline	
		\\[-0.9em]
		
Bordeaux & 576.2 & 277 & 2.1 \\
Clermont-Ferrand & 302.6 & 103 & 2.9 \\
Dijon & 232 & 120 & 1.9 \\
Grenoble & 542.5 & 201 & 2.7 \\
Lille & 675.9 & 517 & 1.3 \\
Lyon & 537.5 & 512 & 1 \\
Le Mans & 267.3 & 96 & 2.8 \\
Marseille & 233.4 & 388 & 0.6 \\
Metz & 306 & 113 & 2.7 \\
Montpellier & 438.9 & 160 & 2.7 \\
Nancy & 142.7 & 115 & 1.2 \\
Nantes & 533.7 & 234 & 2.3 \\
Nice & 161.6 & 189 & 0.9 \\
Orléans & 335.4 & 117 & 2.9 \\
Paris & 812.8 & 2837 & 0.3 \\
Rennes & 710.4 & 171 & 4.2 \\
Saint-Etienne & 723.5 & 187 & 3.9 \\
Strasbourg & 339.8 & 195 & 1.7 \\
Toulouse & 461.1 & 254 & 1.8 \\
Tours & 389.3 & 120 & 3.2 \\
		
		\hline          	
	\end{tabular}
\end{table} 

The distribution of traffic volume (addition of download and upload traffic volume) for the 17 selected services has been stored in a three-dimensional matrix $T=(T_{s,i,h})$ representing the traffic volume associated with mobile service $s \in |[1,17]|$ in IRIS $i \in |[1,6906]|$ during a given hour $h \in |[1,96]|$. The traffic volumes associated with the original 100 x 100 m$^2$ grid cells have been assigned to the IRIS based on the fraction of cell surface area intersecting the IRIS. The traffic has then been normalized to obtain a total traffic volume summing to 1 for each IRIS (Equation \ref{norm}). 
\begin{equation}
	\tilde{T}_{s,i,h} = \frac{\displaystyle T_{s,i,h}}{\sum_{k=1}^S \displaystyle  T_{k,i,h}}
	\label{norm}
\end{equation}
where $S=17$ is the number of selected mobile services. 

Finally, we quantified the level of mobile service diversity in each IRIS with the normalized Shannon diversity index \citep{Shannon1948} computed as follow for a given IRIS $i$ during a given hour $h$:
\begin{equation}
	H_{i,h} = -\frac{1}{ln(S)}\sum_{s=1}^S \tilde{T}_{s,i,h}ln(\tilde{T}_{s,i,h})
	\label{H}
\end{equation}

\subsection*{Land use data}

Land use maps for the 20 cities were extracted from the Copernicus Urban Atlas Land Cover/Land Use 2018 products\footnote{\url{https://land.copernicus.eu/en/products/urban-atlas/urban-atlas-2018}}. Urban Atlas 2018 provides high-resolution land use and land cover data with integrated population estimates for 788 Functional Urban Areas (FUA) with more than 50,000 inhabitants in Europe. It contains 27 land use type in total, 17 urban land use types (with the Minimum Mapping Unit (MMU) of 0.25 ha) and 10 rural land use types (with the MMU of 1 ha). Figure \ref{Fig2} shows the code and description of the different land use types, along with an example of their spatial distribution in the city of Bordeaux.

For each IRIS $i$ and land use type $l$, we extracted the fraction $F_ {i,l}$ of the surface area of IRIS $i$ occupied by the land use type $l$. It should be noted that the sum of the fractions of surface area adds up to 1 for each IRIS (i.e. $\sum_{l=1}^{27} F_ {i,l} = 1$).  

\subsection*{Clustering analysis}

Two clustering analyses were conducted separately: one based on the temporal diversity per IRIS $(H_{i,h})_{h \in |[1,96]|}$ (referred to as TD clustering), and the other based on the land use distribution $(F_ {i,l})_{l \in |[1,27]|}$ (referred to as LU clustering). In both cases, an ascending hierarchical clustering was performed using Ward's metric and Euclidean distances as the agglomeration method and dissimilarity metric, respectively \citep{Hastie2009}. The IRIS were clustered based on their Shannon diversity index over time for TD clustering and land use type distribution for LU clustering. 

We analysed the clusters and their relationships, using V-Test values \citep{Lebart2000} to measure the under- or over-representation of each LU cluster in each TD cluster, as shown in Equation \ref{VT}. 
\begin{equation}
	\displaystyle VT_{lh}=\frac{n_{lh}-\frac{n_l n_h}{n}}{\sqrt{\frac{n-n_h}{n-1}\left(1-\frac{n_l}{n}\right)\frac{n_l n_h}{n}}} \label{VT}
\end{equation}
Where $n$ represents the number of IRIS, $n_l$ the number of IRIS in the considered LU cluster and $n_h$ the number of IRIS in the considered TD cluster. The VT value compares the actual number of IRIS $n_{lh}$ belonging to both clusters at the same time, with the average number $n_h n_l/n$ that would be expected if the LU cluster were uniformly distributed over the whole studied area. Since this quantity depends on $n_l$ and $n_h$ it is normalized by the standard deviation associated with the average expected number of IRIS in cluster TD \citep{Lebart2000}. 

\begin{figure*}
	\centering 
	\includegraphics[width=\linewidth]{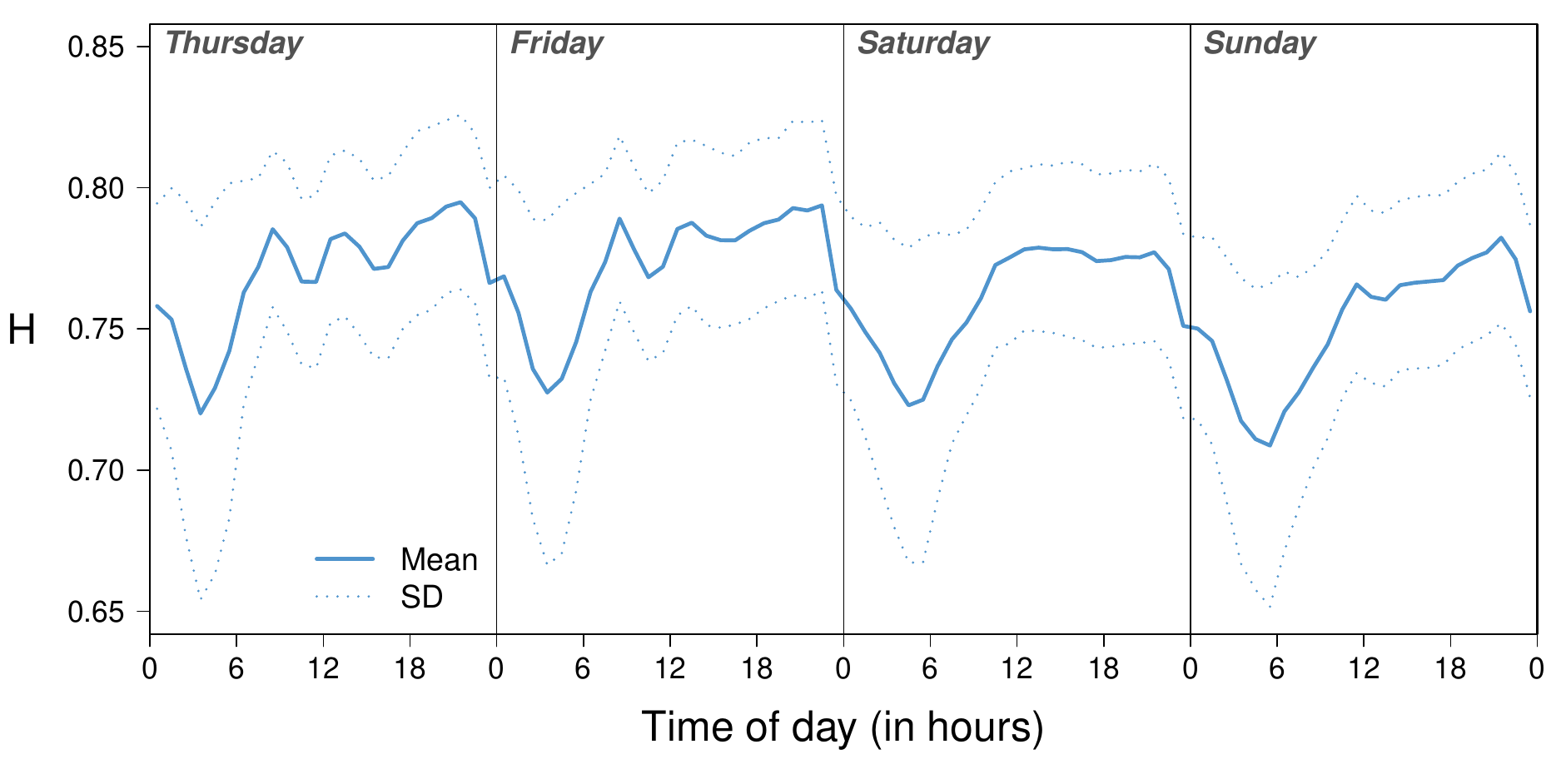}
	\caption{\textbf{Shannon diversity index as a function of time.} The results are based on the average Shannon diversity index for a given hour across all IRIS. The solid line represents the average, while the dotted lines represent the standard deviation.}
	\label{Fig3}
\end{figure*}

\section*{Results}

\subsection*{Global analysis}

We first investigated the time evolution of the Shannon diversity (Figure \ref{Fig3}). We focused here on the average Shannon diversity index per IRIS for a given hour and the associated standard deviation. The diversity is observed to be generally high, fluctuating around 0.763 with an average standard deviation of 0.036. The diversity is slightly higher during weekdays ($\sim$0.77) than during weekend days ($\sim$0.756), and it is lower during the night than during the day. Peaks of diversity are observed around 8-9am, 12pm, and in the evening. These peaks are less pronounced during weekend days and the first two peaks even merged on Saturday.

\begin{figure}[!h]
	\centering 
	\includegraphics[width=\linewidth]{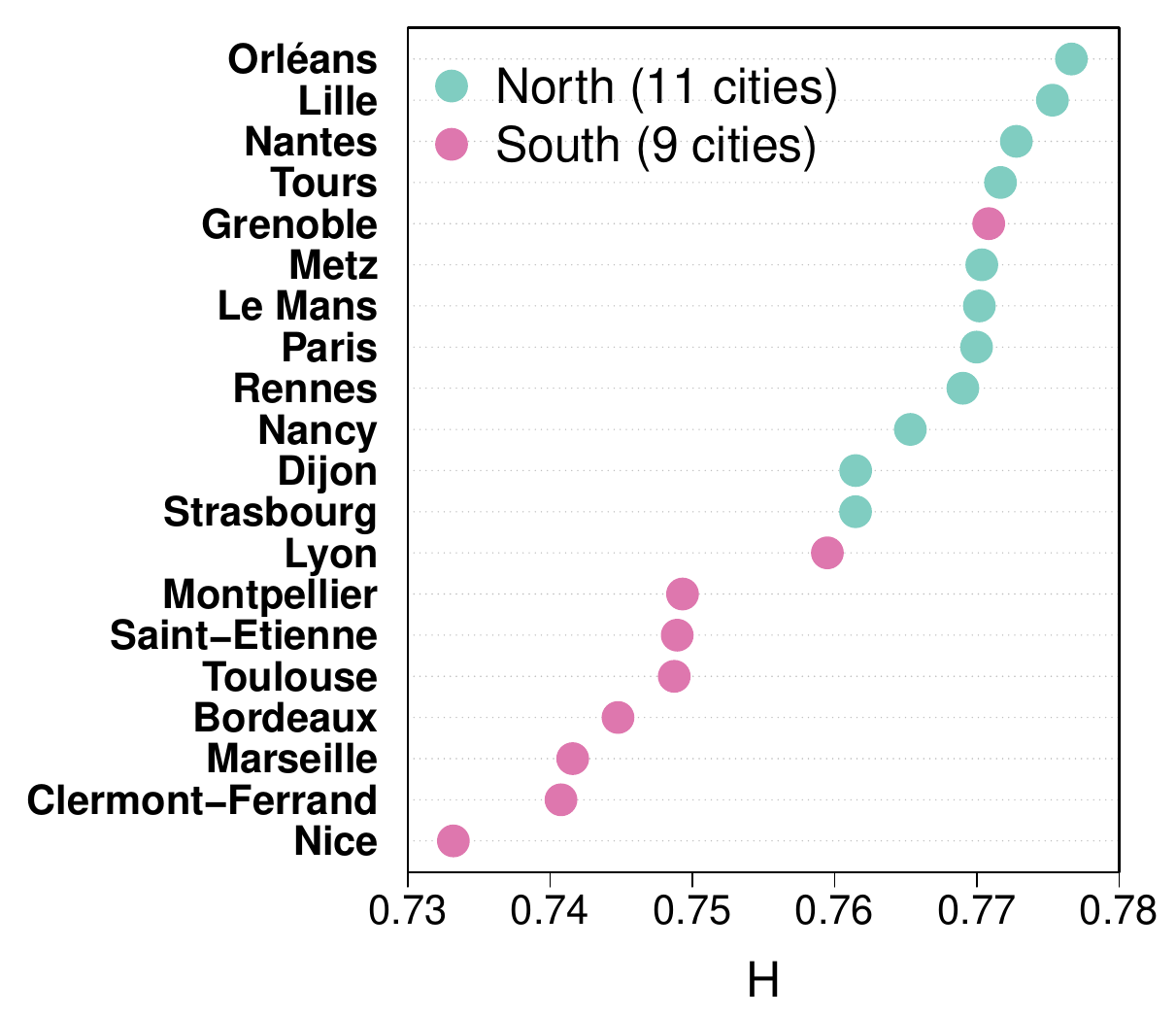}
	\caption{\textbf{Ranking of cities according to the average Shannon diversity.} The results are based on the average Shannon diversity index per IRIS and hour. The cities were differentiated based on their location, either northern (in green) or southern (in red).}
	\label{Fig4}
\end{figure}

\begin{figure*}
	\centering 
	\includegraphics[width=\linewidth]{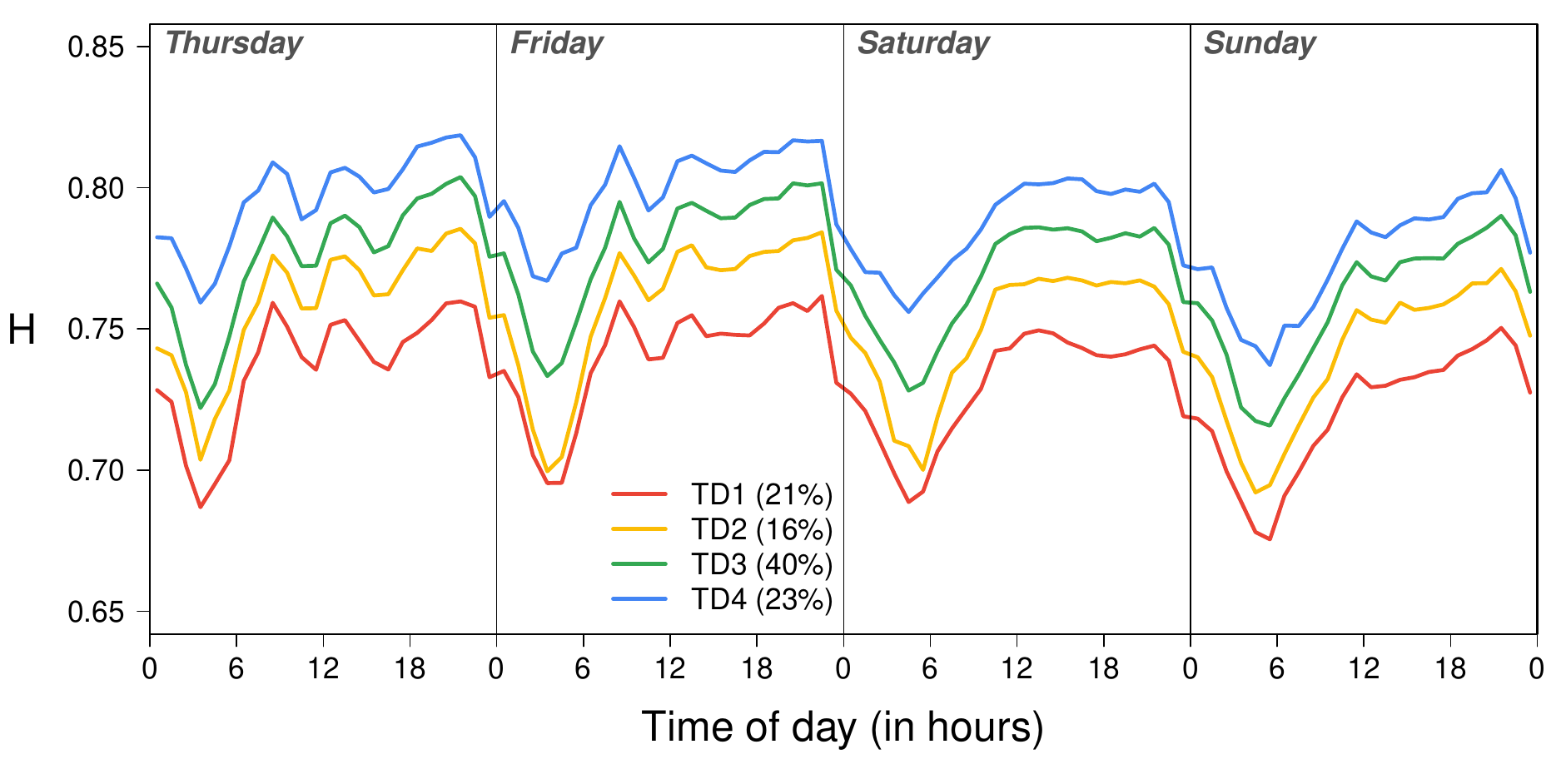}
	\caption{\textbf{Average Shannon diversity index as a function of time by TD cluster.}} 
	\label{Fig5}
\end{figure*}

The 20 cities were ranked according to the Shannon diversity index averaged over IRIS and time, as shown in Figure \ref{Fig4}. A gradient of diversity can be observed among cities based on their locations, with some cities such as Nice, Clermont-Ferrand, Marseille, Bordeaux, Toulouse, Saint-Etienne, and Montpellier having \enquote{low} diversity, while others such as Orl{\'e}ans, Lille, Nantes, Tours, Grenoble, Metz, Le Mans, Paris, Rennes, Nancy, Dijon, Strasbourg, and Lyon having a \enquote{high} average Shannon diversity index. Remarkably, cities belonging to the \enquote{low} diversity group are located in the southern half of France, with the exception of Grenoble and to a lesser extent Lyon (Figure \ref{Fig1}). This ranking is based on the average diversity by IRIS and hour. However, it is important to note that this ranking remains valid when comparing the diversity in cities for each hour with statistical tests (more details are available in Figure S2 in the Appendix).

\subsection*{Clustering analysis}

The clustering algorithm identified four TD clusters and four LU clusters. In both cases, the number of clusters was determined by comparing the ratio between the within-group variance and the total variance (more details in Figure S3 and S4 in Appendix). 

The IRIS have been grouped into four different TD clusters based on their similarity in temporal diversity distribution. Figure \ref{Fig5} displays the average Shannon diversity index over time by TD cluster. The different profiles are similar in shape, following a similar trend to that observed when considering all IRIS. The varying levels of diversity throughout the week appear to be the key factor in differentiating the clusters. The first cluster (TD1) comprises 21\% of IRIS with a diversity fluctuating around 0.73. The second cluster (TD2) is smaller, representing 16\% of IRIS with an average diversity of 0.75. The third cluster (TD3), is the largest, comprising 40\% of IRIS with a diversity fluctuating around 0.77 over the course of the week. Finally, the fourth cluster (TD4) represents 23\% of the IRIS with the highest level of diversity (0.79). In all cases, the average standard deviation is 0.03.

\begin{figure*}
	\centering 
	\includegraphics[width=\linewidth]{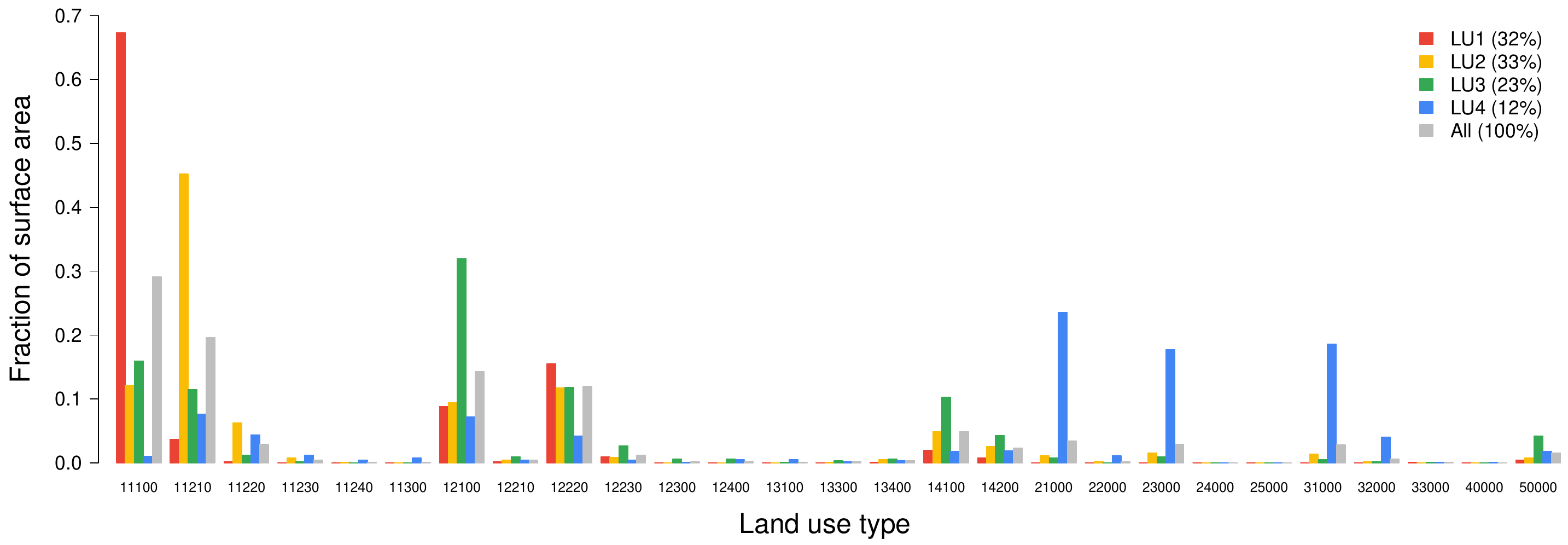}
	\caption{\textbf{Average fraction of surface area by land use type, both for each LU cluster and in total.} The standard deviation can be found in Table S1 in the Appendix.} 
	\label{Fig6}
\end{figure*}

The IRIS have been gathered into four different LU clusters based on their similarity in land use distribution. Figure \ref{Fig6} shows the fraction of surface area by land use type for each LU clusters and all IRIS combined. The first cluster gathering 32\% of the IRIS is a dense urban cluster characterized by an over-representation of \enquote{Continuous urban fabric} (11100) and \enquote{Others roads and associated lands} (12220). The second cluster is of comparable size (33\%). It is over-represented by \enquote{Discontinuous dense urban fabric} (11210) and \enquote{Discontinuous medium density urban fabric} (11220). The LU3 cluster is smaller (23\% of IRIS) and it presents an over-representation of \enquote{Industrial, commercial, public, military and private units} (12100), \enquote{Fast transit road and associated land} (12100), \enquote{Railways and associated land} (12230), \enquote{Green urban areas} (14100), \enquote{Sports and leisure facilities} (14200) and \enquote{Water} (50000). The last cluster is the smallest one, gathering 12\% of the IRIS. It is characterized by an over-representation of \enquote{Discontinuous medium, low and very low density urban fabric} (11220, 11230 and 11240) and \enquote{Isolated structures} (11300), but also \enquote{Arable land (annual crops)} (21000), \enquote{Permanent crop} (22000), \enquote{Pastures} (23000),  \enquote{Forests} (31000) and \enquote{Herbaceous vegetation associations} (32000).

\begin{table}[!h]
	\caption{\textbf{V-Test values quantifying the under- or over-representation of each LU cluster in each TD cluster.}}
	\label{Tab2}
	\centering
	\vspace{0.2cm}
	\begin{tabular}{ccccc}	
		
		\hline
		\\[-0.9em]
		\textbf{Clusters} & \textbf{TD1} & \textbf{TD2} & \textbf{TD3} & \textbf{TD4} \\ 
		\\[-0.9em]		
		\hline	
		\\[-0.9em]
		
		\textbf{LU1} & \textbf{20.37} & \textbf{2.18} & -10.25 & -9.77 \\
		\textbf{LU2} & -6.71 & \textbf{3.71} & \textbf{3.51} & -0.84 \\
		\textbf{LU3} & -4.55 & 0.18 & \textbf{2.54} & 1.3 \\
		\textbf{LU4} & -13.44 & -8.67 & \textbf{6.24} & \textbf{13.43} \\
		
		\hline          	
	\end{tabular}
\end{table}

\begin{figure*}
	\centering 
	\includegraphics[width=\linewidth]{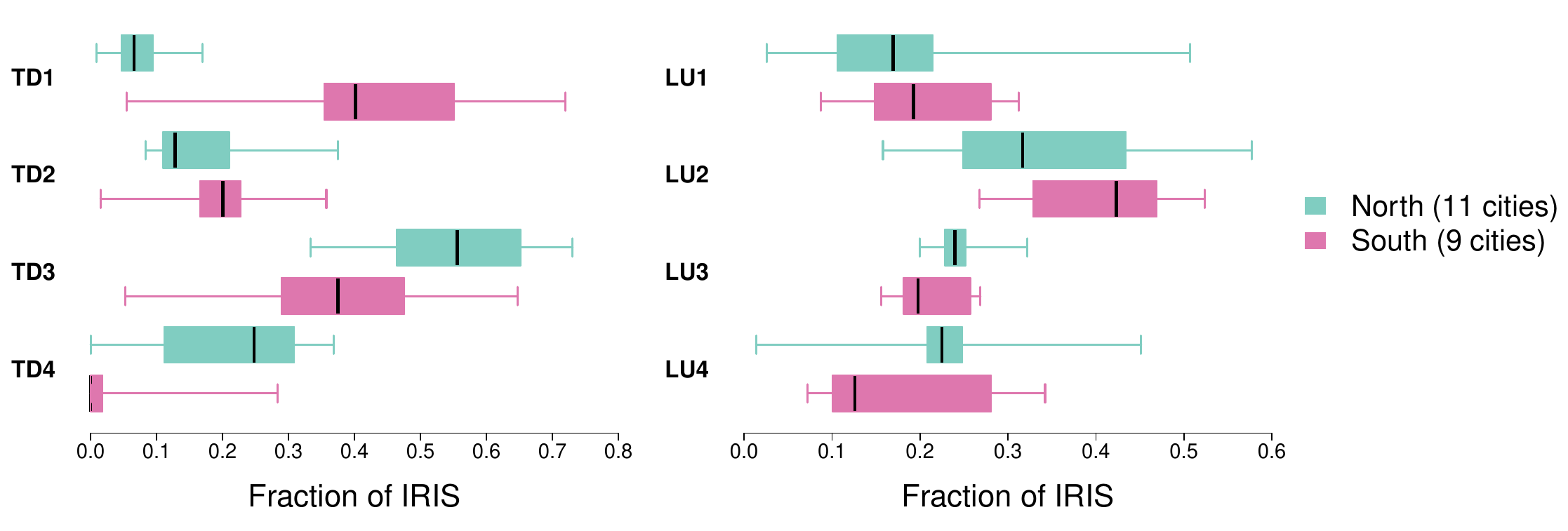}
	\caption{\textbf{Boxplots of the fraction of IRIS belonging to the TD clusters (left) and LU clusters (right) per city.} Each boxplot is composed of the first decile, the first quartile, the median, the third quartile and the ninth decile.} 
	\label{Fig7}
\end{figure*}

We now need to evaluate and understand the relationship between clusters of IRIS based on mobile service usage diversity and land use distribution. To achieved this, we performed a chi-square test to assess the dependence between TD and LU partitions. The hypothesis of no relationship between the two partitions has been rejected with a p-value close to 0 ($< 2.2 \times 10^{-16} $), indicating a significant relationship between the two partitions. Table \ref{Tab2} presents the V-Test values that evaluate the over- or under-representation of IRIS belonging to the LU clusters in the TD clusters. There is an interesting correlation between the level of diversity and the degree of urbanization, ranging from artificial to agricultural and natural landscapes. Specifically, there is a tendency for mobile phone usage diversity to decrease as the level of artificialization increases.

In particular, we observe an over-representation of IRIS belonging to cluster LU1 in cluster TD1 with a positive VT value close to 20.This indicates that the IRIS with lower temporal diversity contain a significantly higher proportion of artificial areas, such as roads and continuous urban fabric, than expected. TD2 is also more characterized by roads and continuous urban fabric (LU1), as well as  discontinuous dense and medium dense urban fabric (LU2), with VT values of 2.18 and 3.71, respectively. The TD3 cluster, which comprises 40\% of IRIS, is also over-represented by IRIS belonging to the LU2 cluster (VT=3.51) but also by the LU3 cluster, which includes land use types dedicated to industrial, logistic and leisure activities (VT=2.54), as well as agricultural areas (VT=6.24). Finally, the fourth TD cluster is characterized by an over-representation of IRIS belonging to the LU4 cluster (VT=13.43).

To evaluate the relationship between diversity and land use as a potential explanation for the observed disparities among cities in Figure \ref{Fig4}, we present the proportion of IRIS assigned to the TD and LU clusters, categorized by city type (northern or southern), in Figure \ref{Fig7}. The IRIS belonging to TD1 and TD2 are more prevalent in southern cities than in northern cities. Conversely, we observe a lower representation of IRIS belonging to TD3 and TD4 in southern cities. Without surprise, even though the differences are less pronounced, this is also true for the LU clustering with more LU1 and LU2 clusters and less LU3 and LU4 clusters in southern cities than in northern cities. 

\section*{Discussion}

In summary, this study analysed the diversity of mobile service usage and its relation to land use distribution within and between cities at different scales. Mobile phone data are generally not publicly available and depending on the market share they may raise concerns about the representativeness of the sample \citep{Barbosa2018,Vallee2023}. However, there are many ways to freely analyse such data without legal constraints, usually through \enquote{data challenges} \citep{Blondel2015,Barbosa2018}. This is the case of this study, which is based on the data provided as part of the \textit{NetMob 2023 Data Challenge} \citep{Martinez2023}. The dataset provides information on the traffic generated by 68 mobile services, at a high spatial resolution of 100 x 100 m$^2$ in 20 metropolitan areas in France during 77 consecutive days in 2019 \citep{Martinez2023}. The unusual richness of this dataset rises interesting questions related to the diversity of mobile service usage in cities. The integration of mobile phone data with land use data offers a powerful tool for understanding urban dynamics and human activity patterns.

In this context, we have shown that the diversity of mobile service usage, as measured by a Shannon diversity index, is generally high but varies significantly between cities. By working at the IRIS level, our study revealed significant differences in the diversity of mobile service use between cities in the southern and northern halves of France. Cities in the southern half showed less diversity than those in the northern half. We found an interesting relationship between the mobile service usage diversity and the distribution of land use at the IRIS level, which allowed us to explain this finding. Indeed, the level of diversity tends to decrease as urbanisation increases. These findings are consistent with research conducted in\citep{Singh2019}, which identified a relationship between urbanization levels and mobile service usage. In our case, the 11 cities located in the northern part of France tend to be more diverse and less urbanized at the IRIS level, in contrast to the 9 southern cities which show less diversity but a higher level of urbanization at the same level.

Although additional data from various countries and sources could be important for validating our findings, we find that a coherent relationship between mobile phone usage diversity and land use organisation emerges. It is important to keep in mind that these findings may be sensitive to the way cities are defined and delineated, as pointed out by \citep{Cottineau2019} while comparing mobile phone and socioeconomic indicators in France. We decided to focus on 17 mobile services among the 68 available in the dataset. It would be important to evaluate the impact of the number and type of mobile services on the findings. Finally, it would be interesting to investigate the relationship between mobile phone usage diversity and socio-demographic and socio-economic features in future studies, as has been done with several mobile phone datasets \citep{Gauvin2020,Lenormand2020,Lenormand2023,Goel2023}.

\section*{Data and materials availability} 

The data was made available by Orange within the framework of the Netmob 2023 Challenge. The code used in the present paper are available online (\url{https://gitlab.com/maximelenormand/mobile-service-diversity}). 

\section*{Acknowledgments}

We would like to thank the NetMob 2023 Challenge Organizing Committee for granting us access to the data. This publication has been prepared using European Union's Copernicus Land Monitoring Service information (\url{https://doi.org/10.2909/fb4dffa1-6ceb-4cc0-8372-1ed354c285e6}).

\bibliographystyle{unsrt}
\bibliography{MSD}	

\onecolumngrid
\vspace*{2cm}
\newpage
\onecolumngrid

\makeatletter
\renewcommand{\fnum@figure}{\sf\textbf{\figurename~\textbf{S}\textbf{\thefigure}}}
\renewcommand{\fnum@table}{\sf\textbf{\tablename~\textbf{S}\textbf{\thetable}}}
\makeatother

\setcounter{figure}{0}
\setcounter{table}{0}
\setcounter{equation}{0}

\newpage
\clearpage
\newpage
\section*{Appendix}

\subsection*{Supplementary figures}

\begin{figure}[!h]
	\centering 
	\includegraphics[width=\linewidth]{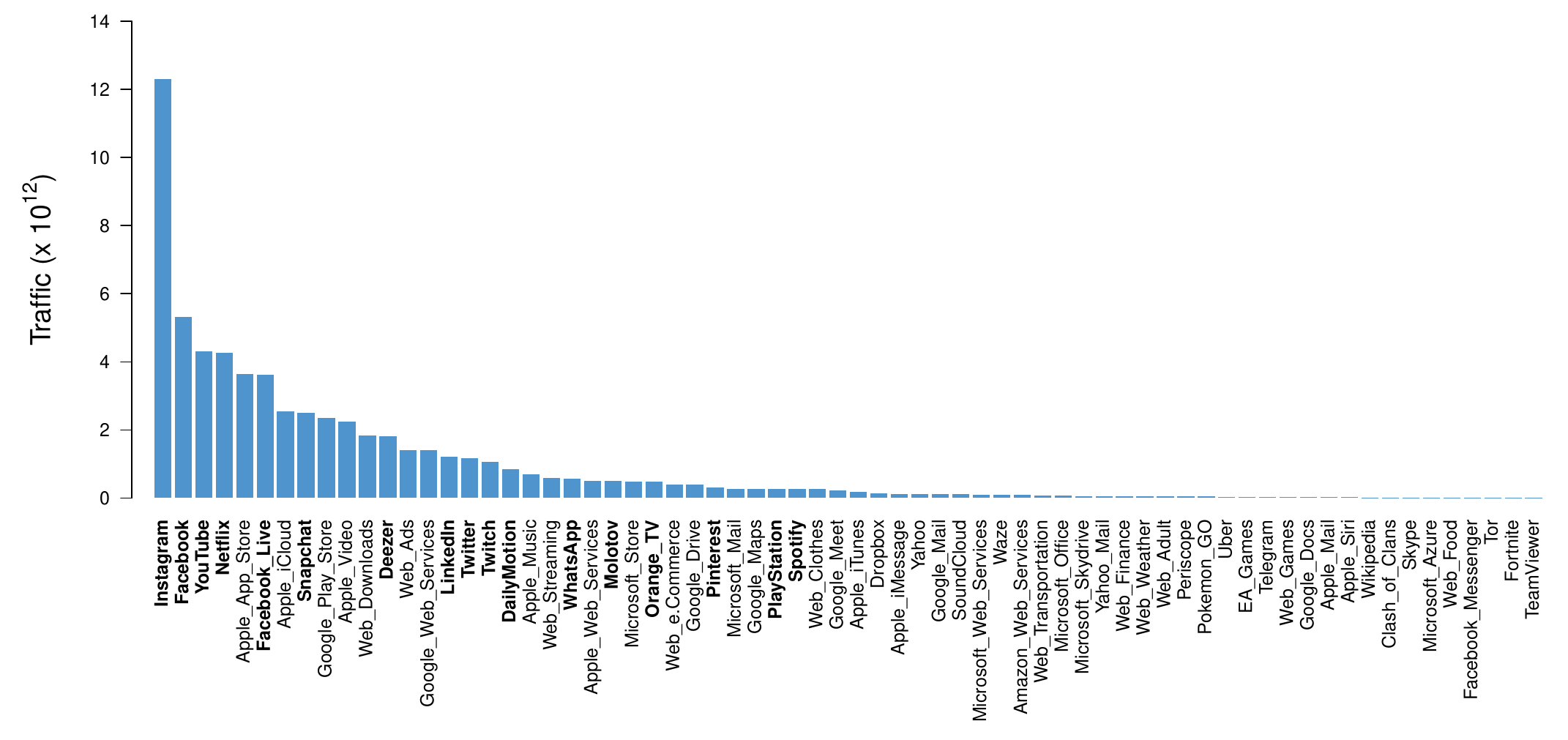}
	\caption{\textbf{Mobile services ranked by the total traffic during a typical week (from Thursday to Sunday) all cities combined.} The 17 services selected for this study are highlighted in bold.} 
	\label{FigS1}
\end{figure}

\begin{figure}[!h]
	\centering 
	\includegraphics[width=13cm]{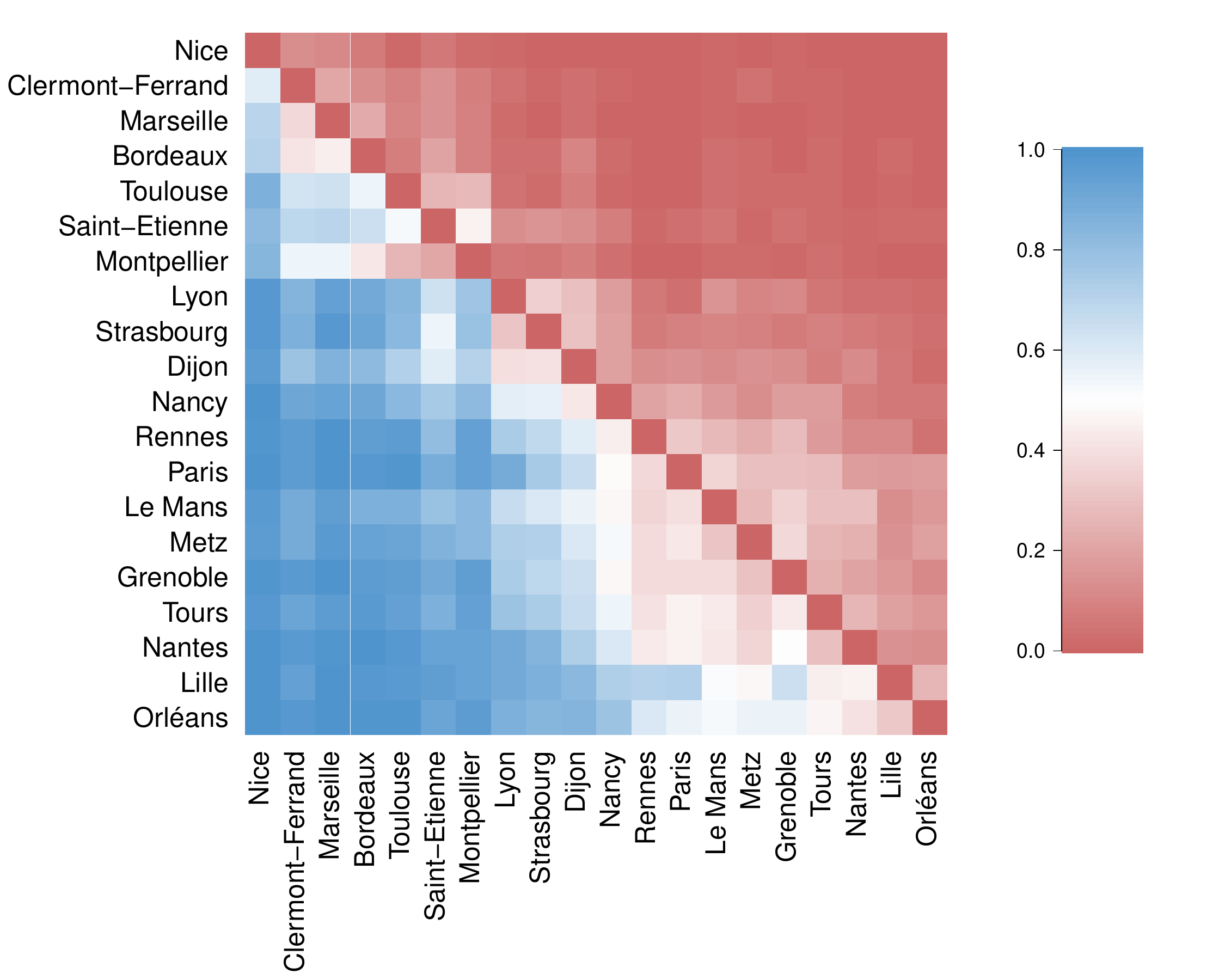}
	\caption{\textbf{Fraction of hours with a median Shannon diversity index (based on the IRIS) significantly greater for the cities in row than for the cities in column.} The fraction of hours is based on the fraction of p-values lower than 0.05 considering a one-sided \enquote{greater} Wilcoxon test.} 
	\label{FigS2}
\end{figure}

\begin{figure}[!h]
	\centering 
	\includegraphics[width=12cm]{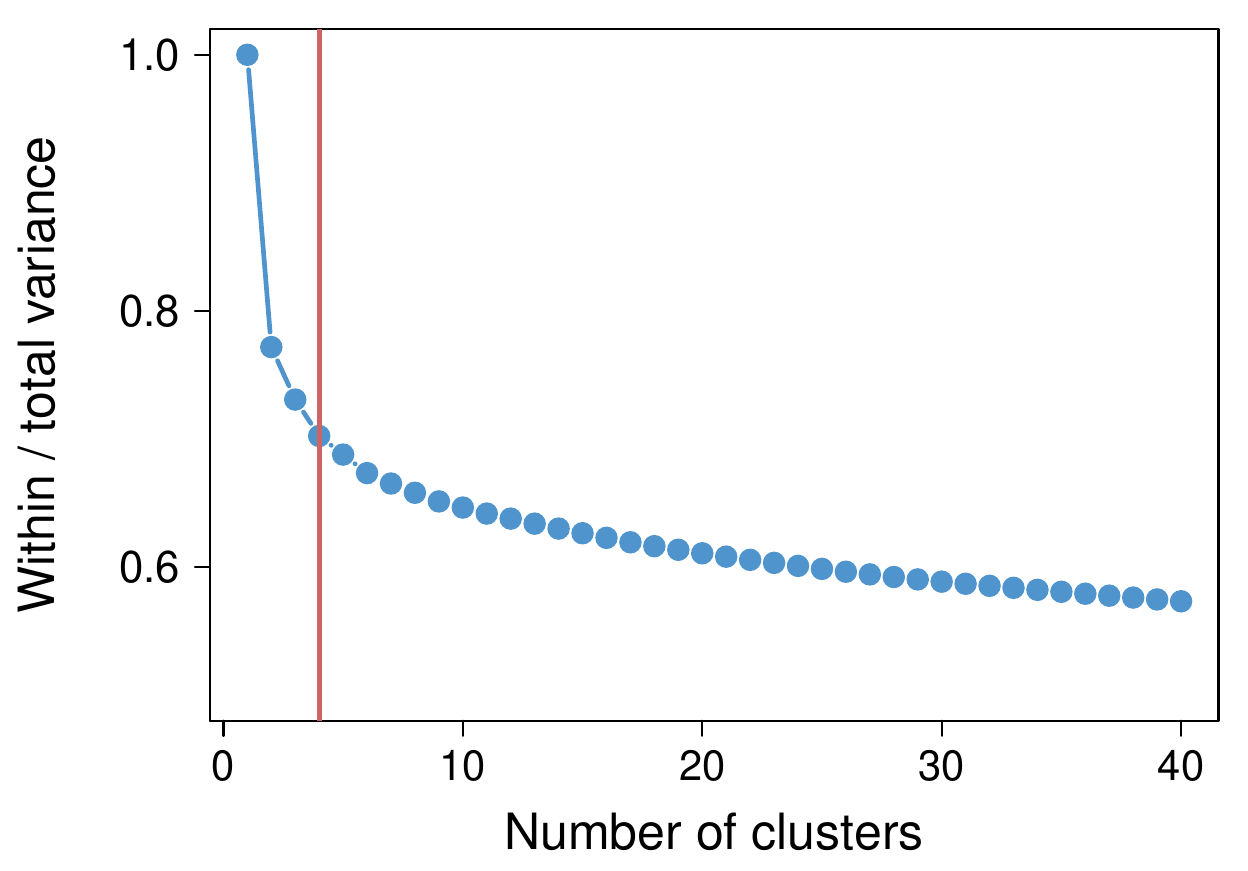}
	\caption{\textbf{Ratio between the within-group variance and the total variance as a function of the number of TD clusters.} We performed an ascending hierarchical clustering using Ward's metric and Euclidean distances as agglomeration method and dissimilarity metric.}
	\label{FigS3}
\end{figure}

\begin{figure}[!h]
	\centering 
	\includegraphics[width=12cm]{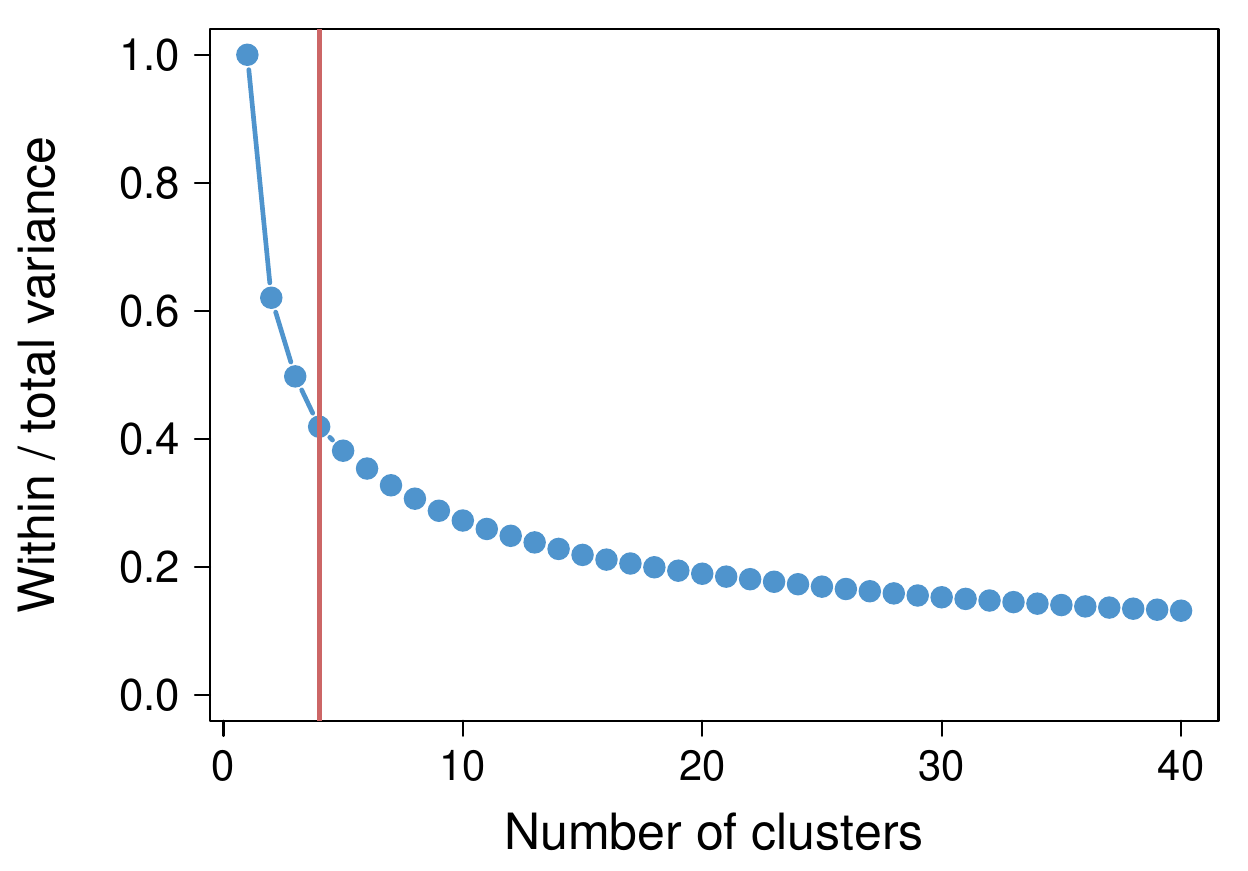}
	\caption{\textbf{Ratio between the within-group variance and the total variance as a function of the number of LU clusters.} We performed an ascending hierarchical clustering using Ward's metric and Euclidean distances as agglomeration method and dissimilarity metric.}
	\label{FigS4}
\end{figure}

\newpage
\clearpage
\newpage
\subsection*{Supplementary table}

\begin{table}[!h]
	\caption{\textbf{Average fraction of surface area per land use type according to the LU clusters and all IRIS combined (standard deviation in brackets).}}
	\label{TabS1}
	\centering
	\vspace{0.2cm}
	\begin{tabular}{lccccc}		
		\hline
		\\[-0.9em]
		\textbf{Type} & \textbf{All} & \textbf{LU1} & \textbf{LU2} & \textbf{LU3} & \textbf{LU4} \\ 		
		\hline	
		\\[-0.9em]
		
11100 & 0.29 (0.29) & 0.67 (0.14) & 0.12 (0.13) & 0.16 (0.14) & 0.01 (0.02) \\
11210 & 0.2 (0.22) & 0.04 (0.07) & 0.45 (0.17) & 0.11 (0.11) & 0.08 (0.07) \\
11220 & 0.03 (0.07) & 0 (0.01) & 0.06 (0.1) & 0.01 (0.03) & 0.04 (0.05) \\
11230 & 0 (0.02) & 0 (0) & 0.01 (0.03) & 0 (0.01) & 0.01 (0.02) \\
11240 & 0 (0.01) & 0 (0) & 0 (0.01) & 0 (0) & 0 (0.01) \\
11300 & 0 (0) & 0 (0) & 0 (0) & 0 (0) & 0.01 (0.01) \\
12100 & 0.14 (0.16) & 0.09 (0.1) & 0.09 (0.08) & 0.32 (0.22) & 0.07 (0.08) \\
12210 & 0 (0.02) & 0 (0.01) & 0 (0.02) & 0.01 (0.03) & 0 (0.01) \\
12220 & 0.12 (0.06) & 0.15 (0.05) & 0.12 (0.04) & 0.12 (0.06) & 0.04 (0.02) \\
12230 & 0.01 (0.05) & 0.01 (0.04) & 0.01 (0.03) & 0.03 (0.09) & 0 (0.01) \\
12300 & 0 (0.02) & 0 (0) & 0 (0) & 0.01 (0.05) & 0 (0.01) \\
12400 & 0 (0.03) & 0 (0.01) & 0 (0) & 0.01 (0.07) & 0.01 (0.04) \\
13100 & 0 (0.01) & 0 (0) & 0 (0) & 0 (0.01) & 0.01 (0.02) \\
13300 & 0 (0.01) & 0 (0) & 0 (0.01) & 0 (0.02) & 0 (0.01) \\
13400 & 0 (0.02) & 0 (0.01) & 0 (0.02) & 0.01 (0.02) & 0 (0.01) \\
14100 & 0.05 (0.09) & 0.02 (0.04) & 0.05 (0.06) & 0.1 (0.15) & 0.02 (0.04) \\
14200 & 0.02 (0.06) & 0.01 (0.03) & 0.03 (0.05) & 0.04 (0.09) & 0.02 (0.03) \\
21000 & 0.03 (0.11) & 0 (0) & 0.01 (0.04) & 0.01 (0.03) & 0.24 (0.2) \\
22000 & 0 (0.02) & 0 (0) & 0 (0.01) & 0 (0) & 0.01 (0.05) \\
23000 & 0.03 (0.09) & 0 (0) & 0.02 (0.05) & 0.01 (0.03) & 0.18 (0.17) \\
24000 & 0 (0) & 0 (0) & 0 (0) & 0 (0) & 0 (0) \\
25000 & 0 (0) & 0 (0) & 0 (0) & 0 (0) & 0 (0) \\
31000 & 0.03 (0.1) & 0 (0) & 0.01 (0.04) & 0.01 (0.02) & 0.19 (0.21) \\
32000 & 0.01 (0.05) & 0 (0) & 0 (0.02) & 0 (0.02) & 0.04 (0.14) \\
33000 & 0 (0.01) & 0 (0.01) & 0 (0) & 0 (0.01) & 0 (0.01) \\
40000 & 0 (0) & 0 (0) & 0 (0) & 0 (0) & 0 (0.01) \\
50000 & 0.02 (0.06) & 0 (0.02) & 0.01 (0.03) & 0.04 (0.11) & 0.02 (0.05) \\
		
		\hline          	
	\end{tabular}
\end{table}

\end{document}